# CMB Anisotropies: An Overview

DOUGLAS SCOTT

*Department of Astronomy and Center for Particle Astrophysics,*
*University of California, Berkeley, CA 94720*

The results from the *COBE* satellite have had an enormous impact on cosmology. The spectral results from FIRAS are impressive, to say the least, and provide strong support for a hot big bang model. However, it seems clear that the day belongs to the DMR measurement of anisotropy[1] and the subsequent flood of other experimental results[2]. Since the *COBE* announcement, there have been roughly 15 reported detections from around 10 separate experiments (see figure). So what progress has been made? How much better do we understand the microwave sky? Can we rule out specific models? What has happened over the last three years — in other words, where are we now and where are we going?

**Progress** There has been genuine progress in both experiment and theory. With the coming of data, theorists and experimenters have actually been talking with one another! One clear indication of progress is that both sets of people are now talking the language of $\ell$-space (where $\ell \sim \theta^{-1}$). The squares of coefficients of spherical harmonics (the $C_\ell$'s) are the natural way to describe a power spectrum on a curved sky. And they are quantities whose expectation values directly come out of theoretical calculations. Perhaps we can finally lay the correlation function to rest.

**Precision** Theorists are now becoming much more precise in their analysis and predictions for experiments. There are many physical effects that should be correctly included. With the potential to measure anisotropies accurately over a range of scales, it is now worth concentrating on understanding $C_\ell$ calculations to at least better than 10%.[3] Among other things it is necessary to properly account for polarization, neutrinos and the physics of recombination

**Normalization** The CMB has also revolutionized another field, namely Large-Scale Structure. Anisotropies on the largest scales are now clearly *the* way to normalize theories. The *COBE* data have enough information about spectral shape that it is not sufficient to use just the rms amplitude, i.e. theories should be normalized to *COBE* in detail. An example of this is the accurate *COBE* normalization of variants of the Cold Dark Matter model[6], e.g. for 'vanilla-flavoured' CDM, $\sigma_8 = 1.34 \pm 0.10$. A point worth stressing, however, is that it is no longer enough to assume that $n \equiv 1$, since there are perfectly reasonable inflationary models with $n \simeq 0.9-0.95$, say. A tilted CDM model with some gravity wave component can have a respectable $\sigma_8 \simeq 0.7$. So before ruling out models, we have to be more careful about the freedom in the initial conditions.

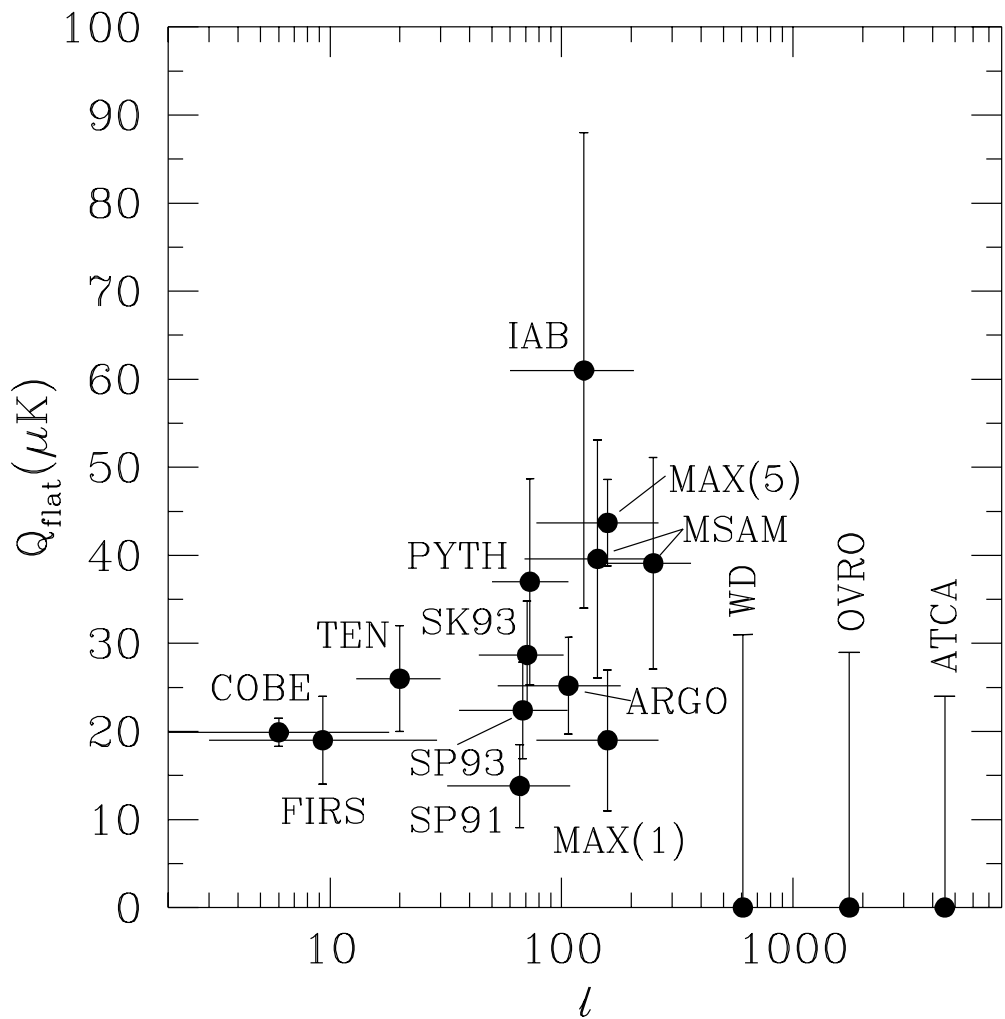

The 'power' in each experiment as a function of scale (multipole $\ell \sim \theta^{-1}$). $Q_{\rm flat}$ is the best-fitting amplitude of a flat power spectrum through the window function of the experiment, quoted at the quadrupole. The vertical error bars are $\pm 1\sigma$, while the horizontal lines represent the half-power ranges of the window functions. References and discussion of the data are presented elsewhere[4,5]. The results of 5 MAX scans have been combined into one point, with the discrepant $\mu$ Peg point plotted separately. The MSAM experiment has two independent modes. The three smaller-scale upper limits are plotted at the 95% confidence level. The general rise in the area around $\ell \simeq 200$ is evidence for a Doppler peak in the radiation power spectrum.

**Consistency** In order to compare experimental results, one crucial task is to translate all the experimental data into a common framework ('$\Delta T/T$' is simply not precise enough any more). This involves accurately calculating the window function for each experiment, then convolving a theoretically predicted curve with the window functions to normalize that theory for each experiment. Since the signal-to-noise for most experiments is still fairly modest, they are only really sensitive to the total 'power' seen through the window, and not to the detailed shape of the $C_\ell$'s. Power through the window is equivalent to the normalization for a theory that is a *flat* power spectrum. Because of this fact, we prefer to translate the results of an experiment into a value $Q_{\text{flat}}$, which is the amplitude of a flat spectrum, quoted at the quadrupole scale for definiteness. This is plotted in the figure for all of the published anisotropy detections[4], together with the newer Saskatoon[7] and ACME/South Pole[8] results. A careful comparison of these $Q_{\text{flat}}$ values shows that, in fact, the experimental results are not 'all over the place', as often stated. It is possible to find a smooth curve that is consistent with all of the results at about the 90% confidence level. Assuming that the data are fairly consistent then, can any conclusions be drawn about cosmology?

**Doppler peaks?** It seems that there is weak evidence for the presence of the so-called Doppler peak at around a degree, i.e. the degree-scale experiments tend to have higher values of $Q_{\text{flat}}$ than the larger angular-scale experiments. No Doppler peak is a worse fit at about the 95% level (assuming Gaussian errors) than the best-fit Doppler peak model[4,5]. This conclusion means that we are starting to learn about the physics of the photon-baryon interactions at redshift $\sim 1000$. The existence of extra power at $\ell$'s of a few hundred is a direct and testable prediction of inflationary models like CDM and its variants.

**Reionization** Even if you are not prepared to take the current evidence for a Doppler peak seriously, it is clear that the Universe cannot have been ionized forever, unless you are prepared to disbelieve the results of several different experiments on degree-scales. This is because such fluctuations would be erased in a fully reionized scenario. Adopting a CDM model with maximal $\Omega_B$ gives an upper limit on the optical depth since reionization of $\tau \lesssim 0.7$. This, I believe, is something new that we have already learned about the Universe from CMB studies.

**$\Omega_0$ & $\Lambda$** What about open or $\Lambda$ models? The position of the main Doppler peak is a very clear indicator of the value of the density parameter $\Omega_0$ — it moves to smaller angular scales for lower $\Omega_0$. It seems that in the next year or two we will have data which will at least allow us to argue about the value of $\Omega_0$ from the CMB. On the largest scales, it is still unclear whether the predictions of open inflationary models[9] will allow a test from *COBE* data. However, it is clear that a constraint on $\Lambda$, competitive with lensing constraints, could be obtained from the 4 years of DMR data[10]. There are also constraints that come from combinations of CMB and galaxy-scale measurements, since the matter amplitude at $z \simeq 0$ and the radiation amplitude at $z \simeq 1000$ have $\Omega_0$ and $\Lambda$ through the growth rate, and the power spectrum shape depends primarily on $\Omega_0 h$.

**Defects** Models with topological defects (e.g. cosmic strings or textures) provide an alternative to inflation for generating fluctuations. Such models may be constrained from *COBE* data alone and generally give non-Gaussian fluctuations at some level. However, it is crucial to obtain detailed calculations of degree-scale fluctuations in non-reionized defect models. Hand-waving arguments indicate that the Doppler peaks may have a testably different shape. There seems little point in discussing such models until the predictions can be confronted with the available data.

**PIB: RIP?** The baryon isocurvature scenario is difficult to rule out because it has a great deal of freedom. However, it seems fair to say that it has a hard time fitting the *COBE* DMR slope and the FIRAS $y$-constraint[11], not to mention the possible Doppler peak. For the CMB, the best-fit model tends to seem contrived to look as much like CDM as possible. But it is still sufficiently different that, as the degree-scale situation improves, it should provide the definitive test.

**Satellites** If there were to be a repeat of the efforts of the *COBE* satellite with today's technology and a significantly smaller angular scale, then we could gather a huge amount of information about the Universe. It would be possible to simultaneously measure $\Omega_0$, $\Omega_B$, $H_0$ and $n$, as well as constraining the cosmological constant, the contribution from gravity waves and the amount of reionization. This is the goal of a number of satellite projects currently being discussed (FIRE, PSI, MAP, COBRAS/SAMBA, ...). If such a project is successful and has enough angular resolution and frequency coverage (to subtract foreground contamination), then many of today's outstanding cosmological questions ought to be answerable.

**Small scales** Fluctuations at the smallest angular scales are best observed from the ground, and this is another area where there will be great expansion in the near future. Arc minute and arc second measurements will come into their own with interferometers, SCUBA, etc. looking for Sunyaev-Zel'dovich fluctuations and dust at high $z$. Theorists have still done little here[12], but there is great scope for learning about galaxy and cluster formation in the mm waveband.

The next few years will be very interesting for getting cosmological parameters from CMB anisotropy studies. Most experimental groups that have reported results have about the same amount of data again, awaiting full analysis. In only a year or two we are likely to be arguing about the question of $\Omega_0$, as well as constraining isocurvature baryon models and defect models for structure formation. And the more distant future looks astonishingly promising. The CMB anisotropy field is currently in a phase of rapid progress (so much so that these comments are likely to be out of date by the time they see print!).

**Acknowledgements** Most of these comments are based on work or discussions with my collaborators Martin White, Wayne Hu, Ted Bunn, Naoshi Sugiyama and Joe Silk.